# Comparing Spatial Navigation and Human–environment Interaction in Virtual Reality vs. Identical Real Environments across the Adult Lifespan

Saleh Kalantari [1*], Bill Tong Xu[1], Armin Mostafavi[1], Anne Seoyoung Lee[1], Qi Yang[1]

[1] Human Centered Design, Cornell University
*Corresponding Author: Saleh Kalantari <sk3268@cornell.edu>

## Abstract

Virtual reality (VR) is increasingly being used as a research platform for investigating human responses to environmental variables. While VR provides tremendous advantages in terms of variable isolation and manipulation, and ease of data-collection, some researchers have expressed concerns about the ecological validity of VR-based findings. In the current study we replicated a real-world, multi-level educational facility in VR, and compared data collected in the VR and real-world environments as participants (n=36) completed identical wayfinding tasks. We found significant differences in all of the measures used, including distance covered, number of mistakes made, time for task completion, spatial memory, extent of backtracking, observation of directional signs, perceived uncertainty levels, perceived cognitive workload, and perceived task difficulty. We also analyzed potential age-related effects to look for heightened VR/real response discrepancies among older adult participants (>55 years) compared to younger adults. This analysis yielded no significant effects of age. Finally, we examined the spatial distribution of self-reported wayfinding uncertainty across the building floorplan, finding that areas in which uncertainty was most pronounced were similar between the real-world and VR settings. Thus, participants appeared to be responding to the same environmental features in the real and VR conditions, but the extent of these responses was significantly different. Overall, the findings suggest that when VR is used to contrast varying environmental design conditions the resulting data should be interpreted cautiously and should not be generalized into real-world conclusions without further validation.

## 1. Introduction

Immersive virtual reality (VR) technologies, which have rapidly advanced over the past two decades, are profoundly reshaping the methods used in human behavior studies (Brookes et al., 2020; Tarr & Warren, 2002). This paradigm shift is a result of VR's realism, cost-effectiveness, and variable-isolation capabilities, which make it tremendously well-suited for laboratory studies (Bohil et al., 2011; Chamilothori et al., 2019; Darfler et al., 2022; Diersch & Wolbers, 2019; Kalantari, 2016; Pan & Hamilton, 2018; Slater & Sanchez-Vives, 2016). One of the key advantages conferred by VR is its capacity to meticulously control experimental settings, allowing for precise manipulations and a high degree of standardization. This rigor makes VR an attractive alternative to real-world environmental psychology research, which often grapples with the challenge of maintaining consistent experimental conditions.

In addition to its role in fundamental research, VR is now commonly used in many types of applied science, such as user testing for novel architectural or product designs. Particularly in the architectural design of large and complex facilities, VR research paradigms make it possible to evaluate issues such as navigational challenges and user–building interactions in an efficient fashion prior to the tremendous financial and logistical investment in facility construction. The VR method has the potential to significantly contribute to evidence-based design practice, making user research more accessible and thereby advancing overall knowledge about human–environment interactions as well as enhancing the performance of specific facilities (Kalantari, Tripathi, et al., 2022; Portman et al., 2015; Whyte & Nikolic, 2018). Such transformative applications of VR in research and design have also extended to the domains of consumer behavior, ergonomics, and human–robot interaction, among others (Verhulst et al., 2017; Whitman et al., 2004; Williams et al., 2018).

While VR provides an extremely effective research platform, its increasing use has prompted concerns about the ecological validity of VR-based studies; that is, the ability to generalize their findings to the real world (Bishop & Rohrmann, 2003; Haq et al., 2005; Kort et al., 2003; Skorupka, 2009; Westerdahl et al., 2006). This has been particularly contentious in the area of spatial navigation research, where VR has made it possible to exert fine-grained control over environmental variables while collecting robust behavioral and physiological data that would be difficult to obtain in real-world contexts, such as "sign-seen" events and neurological responses to environmental features (Claessen et al., 2016; Doeller et al., 2008; Rounds et al., 2020; Sutton et al., 2010; van der Ham et al., 2015). Prior studies directly comparing VR vs. real environments in terms of human perceptions and behaviors related to navigation have generated inconclusive findings. Some research has supported the efficacy of VR-based approaches by finding similar human responses between VR and real contexts (Kuliga et al., 2015; Lloyd et al., 2009; Marín-Morales et al., 2019; Skorupk, 2009), but other studies have found significant discrepancies (Haq et al., 2005; Kimura et al., 2017; S. Kuliga et al., 2020; Savino et al., 2019).

In reviewing the relevant literature, we noted that evidence about VR/real comparisons in spatial navigation research has also been largely confined to healthy young adults, a demographic that may exhibit a learning-curve bias and greater acceptance of VR technology compared to older adults (Cushman et al., 2008). This gap in the research is highly salient given the substantial effects of aging on human sensory and cognitive abilities that are relevant to wayfinding (Harris & Wolbers, 2012; Head & Isom, 2010). It is also of particular concern because one of the advantages of VR-based research is in facilitating studies with broader participant samples and more inclusive demographics (Kalantari & Neo, 2020). If VR research in this area is to fulfill its potential for expanding the range of environmental psychology knowledge and for nudging designers to consider how diverse users interact with the built environment, then it is vital to confirm that the responses collected from such diverse users in VR will mirror how these individuals interact with the features of actual real-world facilities.

The present study was conducted to contribute to the literature on VR/real comparisons in spatial navigation research, and in particular to conduct such comparisons for participants of different adult ages. The research used a between-subject approach with a 2x2 design, including independent variables on the axis of real vs. identical VR, and on the axis of young adults (18–30 years of age) vs. older adults (>55 years of age). We evaluated multiple metrics related to spatial navigation, including standard performance measures (task-completion time, number of wrong turns, etc.) along with novel self-reported measures such as perceived wayfinding uncertainty and task-load.

### 1.1. Previous Human–Environment Interaction Studies Comparing Real vs. VR

Prior work in this area has produced conflicting outcomes, some of which may be attributable to changes in VR technology over time. Sanchez-Vives and Slater (2005), for example, influentially highlighted contrasts between the visual complexity of the real world and that of VR, suggesting that even high-end computer graphics cannot fully replicate the intricate aspects of the physical world in real-time. However, this study was conducted nearly 20 years ago, when the resolution and graphical processing power of the technology was much less advanced compared to today. It should also be noted that the level of immersion and realism produced by VR is not solely dependent on visual fidelity, but also on the quality of user body-tracking, the frame-rate, the latency, and the content design, among other factors (Cummings & Bailenson, 2016). Multiple researchers have since noted that visual fidelity, whether higher or lower, should not be used as a proxy for actual measures of human responses to an environment (Heydarian et al., 2015; S. F. Kuliga et al., 2015; Lloyd et al., 2009b; Paes et al., 2017).

Among studies that have evaluated human-response data, the preponderance of evidence appears to support equivalent outcomes in real and VR environments across a variety of measures. Kalantari and colleagues (2021) evaluated immersive VR environments designed to replicate real-world classrooms, and found no significant differences across multiple physiological measures or in cognitive test accuracy; although they did find that participants responded to questions slightly faster in the VR. Bhagavathula and colleagues (2018) found comparable pedestrian behaviors in terms of intention to cross a street and perceived traffic risk, with a significant VR/real-world difference noted only in the perceived speed of vehicles. In an office setting, Chamilothori and colleagues (2019) observed no significant VR/real-world differences in perceptual accuracy or in physical outcomes such as fatigue. Similarly, Heydarian and colleagues (2015) found comparable performance outcomes in dark and bright conditions in a physical office vs. a virtual office. In the area of consumer research, Siegrist and colleagues (2019) found that participants' choice of cereal products did not statistically differ between virtual and physical stores; and Pizzi and colleagues (Pizzi et al., 2019) reported that VR environments elicited similar utilitarian and hedonic reactions to various consumer products. However, Pizzi and colleagues also noted that participants tended to perceive a reduced product assortment (types of available products) in the VR environment, which negatively affected their reported satisfaction with the retail environment. Most specific to navigational issues, Kuliga and colleagues (2020) reported analogous patterns of wayfinding behavior between identical real and virtual environments, albeit with slightly superior real-world spatial mapping accuracy; and Skorupka (2009) found that participants used similar environmental cues and adopted similar wayfinding strategies in the real-world and in VR.

Despite the positive evidence of VR–real correspondence, researchers have uncovered enough limitations in this area to warrant caution about haphazardly generalizing the findings of VR studies. In addition to the discrepancies noted in the previous paragraph, one of the most notable cautions comes from research conducted by Savino and colleagues (2019), who identified notable differences in VR vs. real-world environments in terms of wayfinding performance and task load. Kuliga and colleagues (2015) also found notable divergences in perceptions of distances and sizes, and in reported perceptions of atmospherics (warmth, attractiveness, invitingness), between VR and real environments. Pontonnier and colleagues (2013) found discrepancies in reported environment-related comfort levels in VR vs. identical real-world environments. These findings suggest that there may be hurdles associated with translating navigation-related results from VR to the real world, which should be unsurprising given that there are inherent differences in the way that VR and real environments are experienced. Real-world movement, in particular, provides a whole-body kinesthetic experience and requires more physical effort than does movement in VR. Virtual systems also tend to fall short, perhaps inevitably so, in replicating the multi-sensory cues inherent in real environments, such as smells, sounds, and fine-grained tactile textures. In some experiments the absence of these cues or, even worse, conflicting cues from the surrounding real environments in which VR experiments take place, may create confounding variables (Ewart & Johnson, 2021). These discrepancies, combined with altered perceptions of speed and distance, and a reduced perceived significance of one's actions, may potentially result in altered behaviors such as diminished head movements and a greater incidence of "tunnel vision" in VR contexts (Ewart & Johnson, 2021).

Finally, it should be noted that there are many different types of VR technology, and that this technology is continuing to develop at a frenetic pace. Thus, correspondences or discrepancies uncovered in a particular study may not continue to be applicable in other studies that use different variants of VR software or hardware. As Ewart and Johnson (2021) have discussed, the analysis of VR/real-world correspondences in the research literature has tended to lag behind the pace of technological evolution, and many conclusions from studies conducted a decade or more ago may no longer be applicable to current VR systems. For this reason, it is vitally important that researchers document the software and hardware aspects of their experimental setup, and continue to pursue ongoing VR/real validation studies of specific technologies and for specific participant measures.

### 1.2. Relevance of Participant Age for Spatial Navigation Studies

The spatial navigation abilities of adults tend to diminish with increasing age—a phenomenon that is linked to the existence of conditions such as Alzheimer's disease in the older adult population, but that also appears to be associated the normal effects of aging on sensory abilities and cognitive skills (Braak & Del Tredici, 2015; Burns, 1999; Lester et al., 2017; Lithfous et al., 2013). These effects of aging make it particularly important for researchers and designers to evaluate how older adults respond to environmental variables, so that their needs can be taken into account in the built environment. VR may provide an effective and safe platform for conducing such studies; and it may also potentially be useful as a tool for obtaining fundamental insights on cognitive aging (Braak & Del Tredici, 2015; Diersch & Wolbers, 2019), and/or enacting therapeutic interventions to help older adults to maintain and enhance their navigational skills (Burns, 1999). For most research applications, however, the usefulness of VR in evaluating the responses of older adults requires assurances that these findings will be generalizable to real-world environments.

There are good reasons to suspect that VR/real discrepancies may be more pronounced in older adult populations than in younger adults. Previous studies have found that older adults tend to be less comfortable with VR environments, for reasons that may range from technological skepticism and reduced familiarity, to heightened vulnerability to "cybersickness," to age-related problems with sensorimotor control and the physical use of VR controllers (Arns & Cerney, 2005; Costello & Bloesch, 2017; Diersch & Wolbers, 2019; Merriman et al., 2018; Schäfer et al., 2006). Researchers have also found that participants with computer gaming experience tend to acclimate more swiftly to navigation and orientation tasks in VR; such gaming experience is much less common among today's older adults than it is among younger adults (Murias et al., 2016; Richardson et al., 2011). For all of these reasons, we believe that it is important to evaluate older adults as a specific population when investigating VR/real-world correspondences in research outcomes. To the best of our knowledge, there are no previous studies at all that have isolated older adults as a demographic when evaluating the ecological validity of navigation-related VR research findings. This is a conspicuous gap in the literature that the current study sought to address.

*1.3. Aims of the Current Study*

**This study entailed several components that are innovative in VR navigation research, including the targeted recruitment of older adults** (>55 years of age) as a specific participant group, and the inclusion of novel wayfinding metrics, such as continuous self-reported navigational uncertainty using a slider device (see Methods). The research was organized around five central questions:

**RQ1: Can significant differences be identified between a real-world environment and an identically designed VR environment in terms of wayfinding performance and spatial learning metrics, including (1a) distance covered during task completion; (1b) number of wayfinding mistakes made; (1c) time required for task completion; and (1d) straight-line directional pointing?**

**RQ2: Can significant differences be identified between a real-world environment and an identically designed VR environment in terms of wayfinding behavioral metrics, including (2a) extent of backtracking; (2b) frequency of noticing directional signs; and (2c) duration of observing directional signs?**

**RQ3: Can significant differences be identified between a real-world environment and an identically designed VR environment in terms of self-reported response factors, including (3a) perceived wayfinding uncertainty; (3b) perceived cognitive workload; and (3c) perceived task difficulty?**

**RQ4: Are any differences observed between the real-world and VR wayfinding outcomes (including performance metrics, behavioral metrics, and self-reported responses) moderated by participants' age?**

**RQ5: Does the spatial distribution of continuous self-reported wayfinding uncertainty differ between the real-world environment and the identically designed VR environment?**

## 2. Methods

The experiment used a 2x2 between-subjects design, with one active independent variable (real-world vs. VR) and one passive independent variable (participant age). A targeted recruitment strategy was used to ensure that participants fell into either the young-adult group (18–30 years of age) or the older adult group (>55 years of age). The assignment between real-world or VR conditions was randomized for each age group. All participants were asked to complete the same 7 wayfinding tasks in a randomized order (see Section 2.3).

### 2.1. Statistical Power and Sample Size

Comparisons between two conditions (real vs. VR) in the form of t-tests at the 0.05 significance level requires 128 samples to reach 0.80 power, assuming a medium effect size (d=0.50). The f-tests used to examine the impact of participant age by condition interaction effects at the 0.05 significance level also requires 128 samples to reach 0.80 power, assuming a medium effect size (f=0.25). Thus, assuming an intra-cluster correlation of 0.15, we needed 34.74 participants to reach an effective sample size of 128, which rounds up to 36 participants total when considering the 2x2 design (9 participants in each of four conditions; effective sample size 132.63).

### 2.2. Participants

We recruited the 36 participants using a targeted convenience sampling method (word-of-mouth and announcements on university e-mail lists). Prior to any research activities, all participants were informed of the study's goals and requirements and signed a consent form. The consent form and all study procedures were approved by the Institutional Review Board (IRB) at

**[deleted for the purpose of blind review]**. Demographic information was collected from participants prior to the experiment and is reported in Table 1. In addition to age and gender,

these descriptive measures include scores on the Spatial Anxiety Scale (Lawton, 1994) and the

Santa Barbara Sense of Direction Scale (SBSOD) (Hegarty et al., 2002). Both of these Likert-

scale instruments have been found to have good internal consistency and have been linked with navigational performance. Other than age, these demographic measures were not treated as variables in the data analysis; and it is important to note that the study did not evaluate other variables that might potentially affect responses to VR, such as ethnicity, income level, and geographic background.

**Table 1.** Participant Information

|  | Older Adult | | Young Adult | | Overall | |
| --- | --- | --- | --- | --- | --- | --- |
|  | **Real** | **VR** | **Real** | **VR** | **Real** | **VR** |
| **Age** | 65.72 (5.84) | 67.22 (3.83) | 22.00 (3.35) | 20.78 (2.49) | 43.86 (22.96) | 44.00 (24.10) |
| **Gender** |  |  |  |  |  |  |
| Women | 5 (56%) | 7 (78%) | 6 (67%) | 5 (56%) | 11 (61%) | 12 (67%) |
| Men | 4 (44%) | 2 (22%) | 3 (33%) | 4 (44%) | 7 (39%) | 6 (33%) |
| **Spatial Anxiety** | 2.27 (0.69) | 2.54 (0.77) | 2.64 (0.93) | 2.58 (0.80) | 2.46 (0.82) | 2.56 (0.76) |
| **Sense of Direction** | 5.30 (0.96) | 4.47 (0.85) | 4.24 (1.20) | 4.18 (1.16) | 4.74 (1.19) | 4.33 (1.00) |

*Note:* Self-reported gender is shown as n and (%). Other measures are shown as mean and (SD). One older adult participant in the real-world condition did not report Spatial Anxiety and Sense of Direction data; this participant was omitted in the measurements of those variables.

*2.3. Environment and Procedure*

**The facility used in the experiment was [deleted for the purpose of blind review]**. This educational facility was selected **due to its distinctive and complex architectural design, which includes two multi-level buildings linked through a central commons area. The buildings are of different ages, and the older building has gone through multiple renovation cycles. The resulting intricate layout presented a variety of challenging navigational tasks for participants. To create the virtual replica of the facility, we made use of architectural documents as well as photographs of the relevant interior spaces. This allowed us to precisely replicate both the dimensions of the layout and its interior features such as furniture placements and directional signs (Figure 1).** Modeling and UV mapping for the virtual environment was conducted using Autodesk 3ds Max. Texturing, lighting, and interactions components were then created using the Unreal Engine 4.26. All of the front-end interaction leveraged the Blueprint platform and C++.

The VR environment was presented to participants using a Meta Quest 2 head-mounted display connected to a desktop gaming computer. The VR display had a resolution of 1832 x 1920 pixels per eye, a horizontal field of view of 90°, a refresh rate of 90 Hz, and allowed customization to accommodate participants with different interpupillary distances. Participants engaged with the VR from a standing position, and used a Meta Quest 2 hand-held controller joystick to move within the environment. They were able to look around the environment by rotating their bodies or heads, but other physical movements beyond the controller did not map to the VR. Wayfinding task assignments and completion notices, as well as post-task self-report questionnaire items, were presented directly within the VR using widgets. The VR platform also allowed us to easily measure some of the wayfinding performance variables, such as distance traveled and task duration.

For the real-world environment, a researcher trailed each participant at a distance of approximately 1.5–3.0 meters during the wayfinding tasks. This researcher carried a Microsoft Surface Pro tablet computer running the **[removed for the purpose of blind review]** App, which was developed by our team to collect wayfinding performance measurements such as distance traveled. The researcher also wore a body camera that filmed the participants' real-world wayfinding activities. This recorded video, and the equivalent screen-capture video from the VR sessions, was later analyzed by the researchers to extract additional performance and behavior measures, as discussed in Section 2.4 below. After the end of each real-world wayfinding task, responses to the self-report questionnaires were collected orally by the researcher (Figure 2).

During the experiment, participants in both the real-world and VR conditions were asked to accomplish the same 7 wayfinding tasks with pre-defined starting and ending points. The 7 tasks were arranged in a continuous loop, with the next task starting where the previous one ended. Each participant was assigned to a random starting task number; that is, they began at different places in the loop. The tasks involved varying degrees of difficulty, and included both single-floor and multi-floor wayfinding challenges (Table 2 and Figure 3).

**Figure 1.** Correspondence between the Real-world Facility and the VR Environment

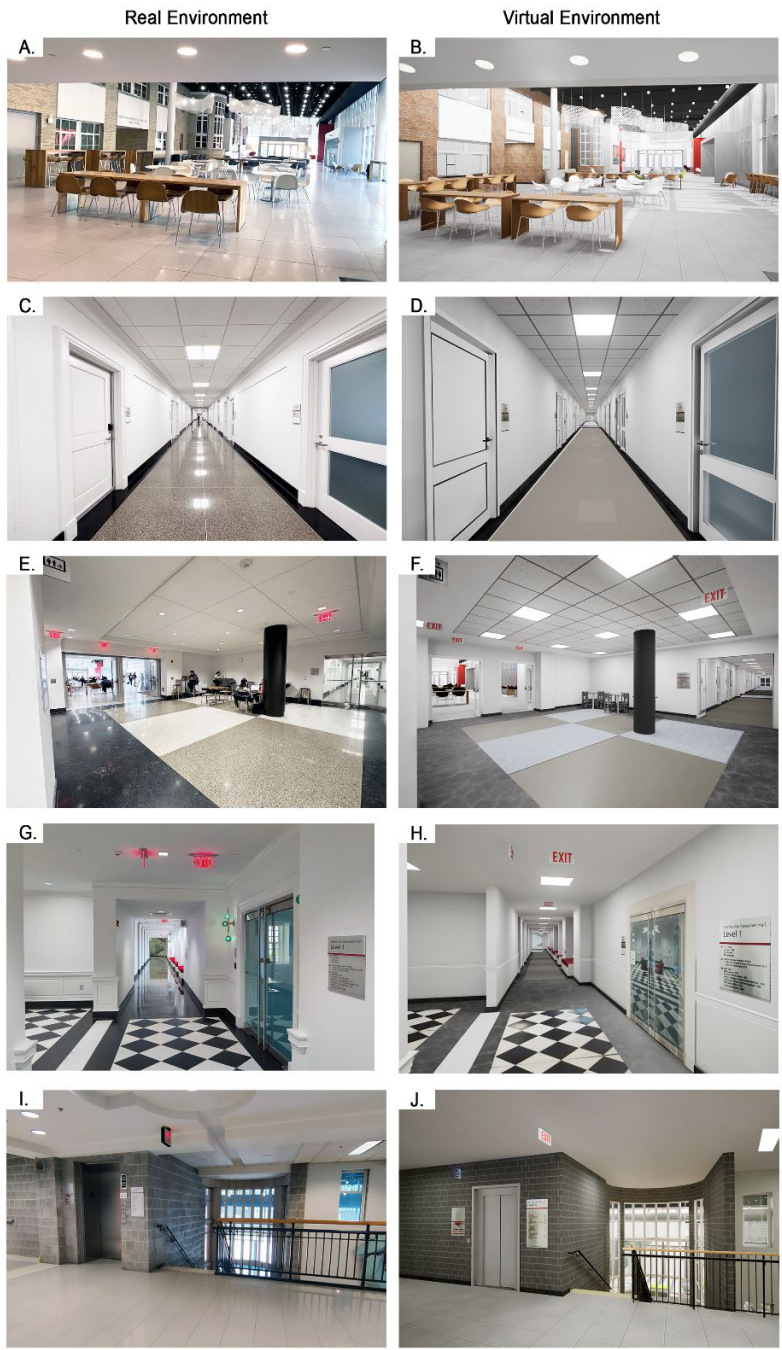

**Figure 2.** Participants in (A) the Real-world Facility and (B) the VR Environment (during Actual Real-world Data-collection the Researcher Trailed the Participant by 1.5–3.0 Meters)

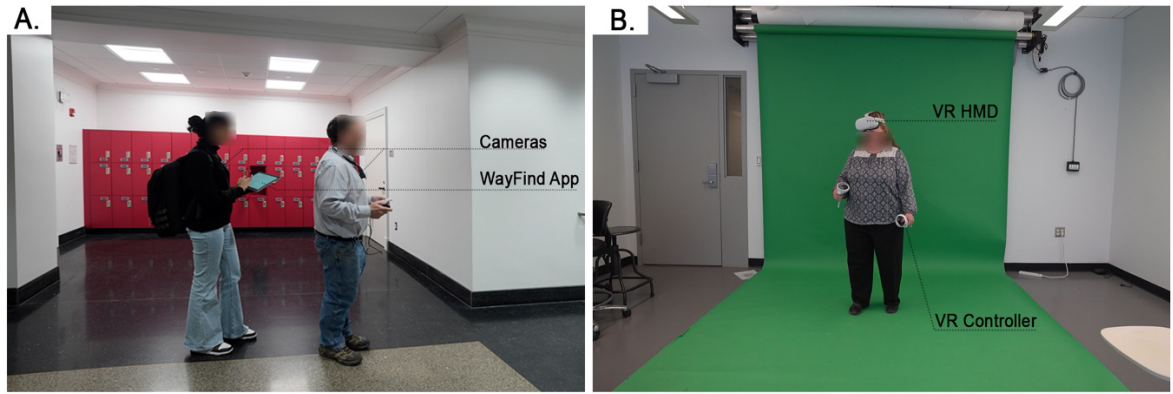

**Table 2.** The Seven Wayfinding Tasks Used in the Study

| Task Number | Start Location | End Location | Change in Floor Levels |
|---|---|---|---|
| 1 | Cafe | Room T222 | Single level |
| 2 | Room T222 | Room T320C | Single level |
| 3 | Room T320C | Room 1250 | Multi-level |
| 4 | Room 1250 | Room 1106 | Single level |
| 5 | Room 1106 | Room G151 | Multi-level |
| 6 | Room G151 | Room HEB T70 | Multi-level |
| 7 | Room HEB T70 | Cafe | Single level |

**Figure 3.** All Wayfinding Tasks as Shown on Building Plans

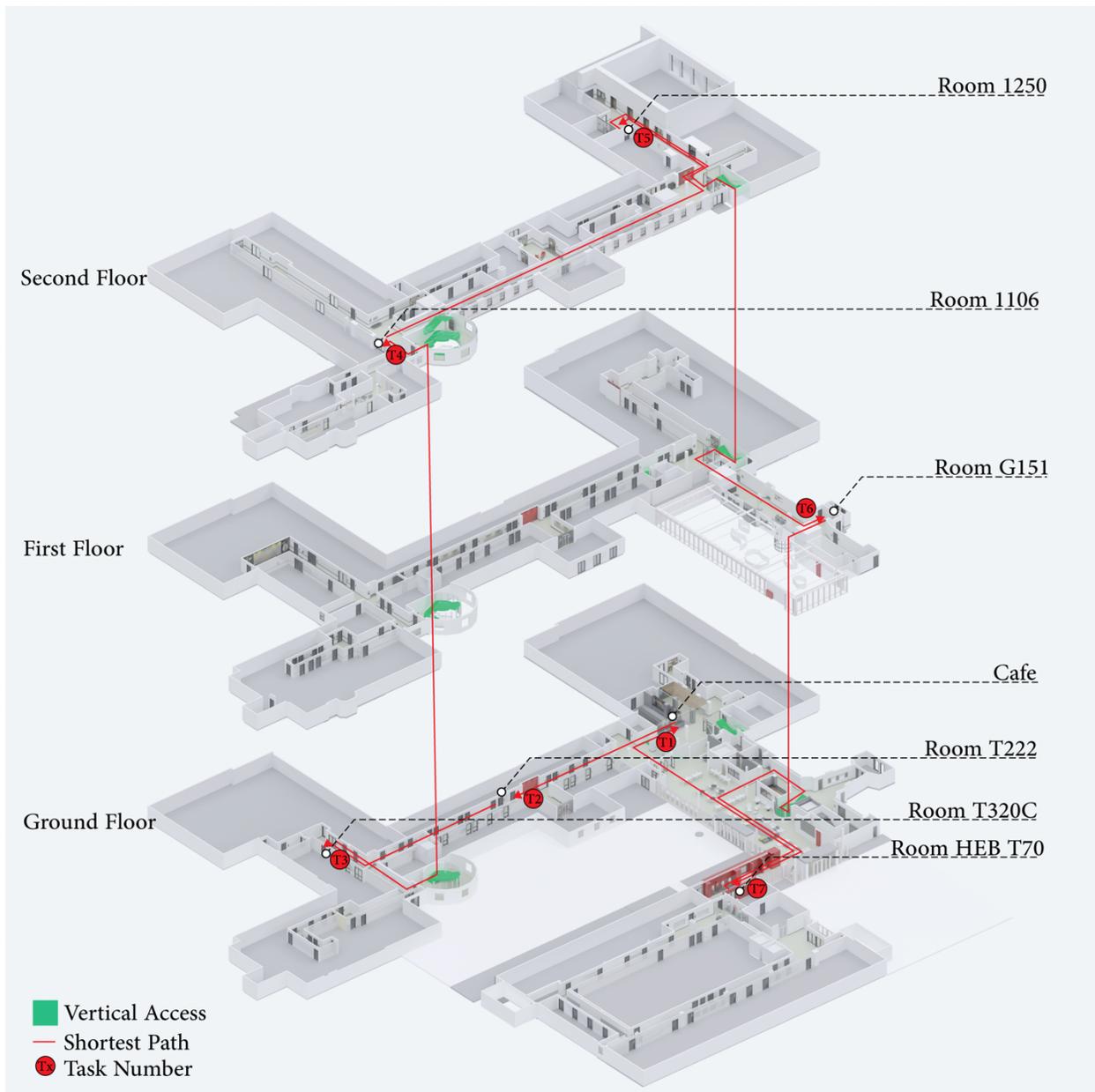

Room 1250

Room 1106

Room G151

Cafe

Room T222

Room T320C

Room HEB T70

Second Floor

First Floor

Ground Floor

■ Vertical Access
— Shortest Path
● Task Number

## 2.4. Measures

Participant trajectories were recorded as a set of timestamps (1 second intervals) associated with floorplan locations. This information, as well as the starting and ending times for each task, was recorded automatically via the **[removed for the purpose of blind review]** App for real-world wayfinding, and through a customized Python script in the VR context. This allowed us to determine the **distance traveled** and the **task duration**.

From the recorded videos of the real-world and VR wayfinding sessions, we extracted several additional metrics. **Number of mistakes** was measured according to wrong turns at decision-points, which were pre-defined on the floorplan as intersections where participants had an opportunity to select among different routes. **Backtracking** was measured as the number of times a participant made a U-turn, including at decision-points. **Sign-seen frequency** was defined as the number of times a participant adjusted their gaze direction to examine an informational sign. **Sign-seen duration** was defined as the total time spent looking at signs during the task, measured from when the participant started to look at an informational sign until they turned away from it. To measure **wayfinding uncertainty,** we played the video recordings for participants post-task and asked them to use a slider device (linear potentiometer) to report their own levels of uncertainty experienced during the task in a continuous fashion. This measure was normalized on a scale from 0 to 1. It allowed us to obtain a record of the distribution of uncertainty across the floorplan, as well as a total task-level uncertainty score which was calculated by adding up the joystick values for each annotation throughout the wayfinding task.

The post-task questionnaires were used to measure participants' perceived cognitive **workload** via the NASA-Task Load Index (Hart & Staveland, 1988); as well as the perceived task **difficulty,** using a single 10-point Likert-scale question: "How difficult did you find this wayfinding task? (1=Extremely Easy; 10=Extremely Hard)." Finally, the post-task questions also included an item to measure **directional pointing**. Standing at the task conclusion point, participants were asked to point in a straight-line horizontal direction toward the task's starting point (vertical direction was ignored for the purpose of this measurement). We measured the angle difference between the indicated direction and the actual direction, using a protractor in the real-world context and the camera angle in the VR context. (The actual angle was determined from the building floorplans.) The difference was divided by 180 degrees, yielding a value between 0 and 1, with smaller values representing lower error.

Finally, after the completion of the experiment we collected two technology response measures only for the VR groups. Participants were asked to rate the **usability** of the VR through the System Usability Scale (Brooke, 1996). A total score on this 10-item, 5-point Likert scale was calculated and normalized to 0–100 range. The VR participants were also asked to complete the Igroup Presence Questionnaire (IPQ) (Schubert et al., 2001) to measure their experience of **spatial presence** ("being there") in the VR environment. This 14-item instrument includes three subscales: Presence, Involvement, and Realism. The original IPQ uses a 7-point Likert scale, but we adjusted this to a 5-point scale for consistency with the other study instruments.

### 2.5. Data Analysis

The R language was used for data analysis. First, we reported descriptive results for each age group and each experimental condition. To evaluate RQ1–RQ3, we used the libraries "lme4" and "lmerTest" to fit linear mixed models with fixed effects of experimental condition (VR/real), age group (young adult/older adult), and their interaction effect; plus the random effect of task and the random effect of participant as control variables. We then used the library "emmeans" to estimate and compare the marginal means of the two experimental conditions, followed by t-tests with the Kenward-Roger method.

For RQ4, we performed f-tests for the age groups by experimental condition interaction effect, using Satterthwaite's method. In addition to the f-test p-values, we also calculated the BIC-estimated Bayes Factors ($BF_{10}$) using the library "bayestestR," which represented the ratio of the posterior probabilities of the predictor ("model 1") to the posterior probabilities of the effect not included ("model 0"). This analysis assumes equal prior probabilities, following the formula $BF_{10} \approx \exp[(BIC_0-BIC_1)/2]$, where $BIC_0$ and $BIC_1$ represent the Bayesian Information Criteria of model 0 and model 1 (Wagenmakers, 2007).

For RQ5 we first divided the floorplan into "zones." Some of these were defined as decision-making zones, following the previous identification of decision-points for the *number of mistakes* analysis. The remaining zones were defined as non-decision, consisting mostly of long corridors, staircases, and dead-ends. We then created visualizations of the spatial distributions of wayfinding uncertainty on the floorplan per experimental condition. We calculated average uncertainty levels per participant per zone, and fitted linear mixed models with fixed effects of experimental condition (VR/real), zone type (decision-making or not), and zone-type interaction effects, plus the random effects of participant and zone as control variables. We then performed f-tests using Satterthwaite's method.

## 3. Results

### 3.1. Descriptive Statistics

Descriptive statistics for all of the study variables are presented in Table 3. On average, participants made 1.47 mistakes per task (SD=2.37), and backtracked 1.10 times per task (SD=1.93). The overall perceived wayfinding uncertainty level (0.31, SD=0.29, range 0–1), perceived  difficulty level (3.60, SD=2.74, range 1–10), and perceived cognitive workload (0.32, SD=0.16, range 0–1) were moderately low. In addition, the VR participants reported a moderately high usability rating for the technology (65.00, SD=25.96, range 0–100) and moderate ratings across all three dimensions of the IPQ (Presence 3.48, SD=0.56, range 1–5; Involvement 2.98, SD=0.63, range 1–5; Realism 2.67, SD=0.66, range 1–5).

**Table 3.** Descriptive Results by Group

| | Older Adult | | Young Adult | | Overall | |
|---|---|---|---|---|---|---|
| | **Real** | **VR** | **Real** | **VR** | **Real** | **VR** |
| Distance Traveled | 136.72 (113.68) | 201.32 (165.14) | 109.24 (112.62) | 155.10 (132.54) | 122.98 (113.54) | 178.82 (151.29) |
| Number of Mistakes | 0.91 (1.60) | 2.47 (2.87) | 0.53 (1.31) | 1.98 (2.82) | 0.72 (1.47) | 2.23 (2.84) |
| Task Duration | 201.43 (223.43) | 268.92 (205.21) | 110.12 (102.45) | 140.34 (119.39) | 155.77 (179.08) | 206.31 (180.19) |
| Backtracking | 0.76 (1.16) | 1.97 (2.50) | 0.25 (0.78) | 1.54 (2.35) | 0.51 (1.02) | 1.76 (2.43) |
| Sign-seen Frequency | 2.54 (3.67) | 17.19 (19.66) | 1.05 (2.32) | 14.07 (18.44) | 1.79 (3.15) | 15.67 (19.06) |
| Sign-seen Duration | 24.22 (45.56) | 29.14 (42.59) | 5.55 (14.25) | 18.64 (26.16) | 14.89 (34.90) | 24.03 (35.79) |
| Uncertainty | 0.27 (0.27) | 0.42 (0.32) | 0.19 (0.23) | 0.36 (0.29) | 0.23 (0.26) | 0.39 (0.31) |
| Workload | 3.30 (2.89) | 5.22 (2.95) | 2.27 (1.73) | 3.73 (2.43) | 2.79 (2.43) | 4.50 (2.80) |
| Difficulty | 0.28 (0.13) | 0.40 (0.16) | 0.28 (0.13) | 0.35 (0.17) | 0.28 (0.13) | 0.38 (0.17) |
| Directional Pointing | 0.25 (0.25) | 0.39 (0.28) | 0.23 (0.25) | 0.35 (0.29) | 0.24 (0.25) | 0.37 (0.28) |

*Note:* All measures are shown as mean and (SD).

## 3.2. Wayfinding Performance and Spatial Learning (RQ1)

The average distance in meters covered during the wayfinding tasks was 122.98 (95% CI 64.63–181.33) for the real-world condition, and 178.77 (95% CI 120.28–237.27) for the VR condition.

The difference in these measurements is 55.79 (95% CI 24.25–87.33), which is significant, t(31.64)=3.60, p=0.001, with a large effect size (Cohen's d = 1.28) (Figure 4a).

The average number of mistakes made during wayfinding tasks was 0.80 (95% CI -0.22–1.83) for the real-world condition, and 2.23 (95% CI 1.20–3.26) for the VR condition. The difference in these measurements is 1.43 (95% CI 0.87–1.97), which is significant, t(31.39)=5.30, p<0.001, with a large effect size (Cohen's d = 1.89) (Figure 4b).

The average wayfinding task duration in seconds was 155.77 (95% CI 93.53–218.02) for the real-world condition, and 205.26 (95% CI 142.66–267.87) for the VR condition. The difference in these measurements is 49.49 (95% CI 4.20–94.78), which is significant, t(31.69)=2.23, p=0.033, with a medium effect size (Cohen's d = 0.79) (Figure 4c).

The average spatial learning measurement based on directional pointing was 0.24 (95% CI 0.14–0.33) for the real-world condition, and 0.37 (95% CI 0.27–0.47) for the VR condition. The difference in these measurements is 0.14 (95% CI 0.04–0.23), which is significant, t(31.90)=2.91, p=0.007, with a large effect size (Cohen's d = 1.03) (Figure 4d). Note that the higher VR scores here indicate a larger error in the pointing task, and thus a *reduced* pointing accuracy.

**Figure 4.** Comparisons of Wayfinding Performance and Spatial Learning Metrics between Real-world and VR Conditions; Error Bars Indicate Model-estimated 95% Confidence Intervals

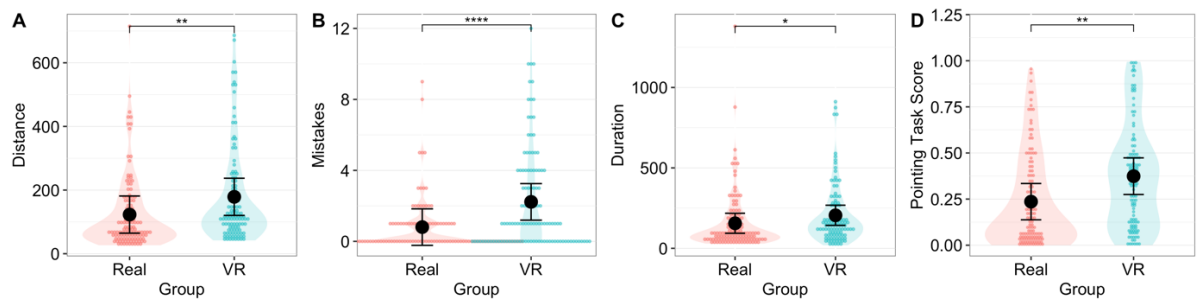

### *3.3 Wayfinding Behavioral Metrics (RQ2)*

The average incidence of backtracking was 0.51 (95% CI -0.23–1.25) for the real-world condition, and 1.75 (95% CI 1.01–2.49) for the VR condition. The difference in these measurements is 1.24 (95% CI 0.80–1.69), which is significant, t(31.64)=5.71, p<0.001, with a large effect size (Cohen's d = 2.03) (Figure 5a).

The average sign-seen frequency was 1.79 (95% CI -1.52–5.11) for the real-world condition, and 15.83 (95% CI 12.44–19.22) for the VR condition. The difference in these measurements is 14.03 (95% CI 9.81–18.26), which is significant, t(31.79)=6.77, p<0.001, with a large effect size (Cohen's d = 2.40) (Figure 5b).

The average sign-seen duration was 14.89 (95% CI 6.16–23.62) for the real-world condition, and 24.21 (95% CI 15.30–33.13) for the VR condition. The difference in these measurements is 9.33 (95% CI -1.43–20.08), which is marginally significant, t(31.78)=1.77, p=0.087, with a medium effect size (Cohen's d = 0.63) (Figure 5c).

**Figure 5.** Comparisons of Wayfinding Behavioral Metrics between Real-world and VR Conditions; Error Bars Indicate Model-estimated 95% Confidence Intervals

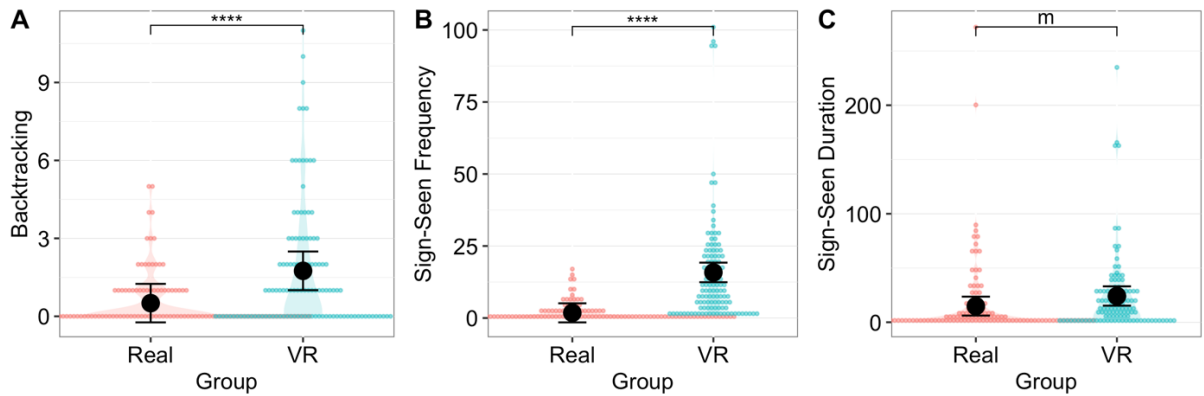

### 3.4. Self-reported Response Metrics (RQ3)

The average uncertainty that participants associated with the wayfinding tasks was 0.23 (95% CI 0.12–0.35) for the real-world condition, and 0.39 (95% CI 0.28–0.51) for the VR condition. The difference in these measurements is 0.16 (95% CI 0.02–0.31), which is significant, t(31.99)=2.25, p=0.032, with a medium effect size (Cohen's d = 0.80) (Figure 6a).

The average task workload score was 0.28 (95% CI 0.22–0.34) for the real-world condition, and 0.38 (95% CI 0.32–0.44) for the VR condition. The difference in these measurements is 0.10 (95% CI 0.04–0.16), which is significant, t(31.93)=3.47, p=0.002, with a large effect size (Cohen's d = 1.23) (Figure 6b).

The average task difficulty rating was 2.79 (95% CI 1.83–3.74) for the real-world condition, and 4.47 (95% CI 3.51–5.44) for the VR condition. The difference in these measurements is 1.69 (95% CI 0.64–2.74), which is significant, t(31.94)=3.27, p=0.003, with a large effect size (Cohen's d = 1.16) (Figure 6c).

**Figure 6.** Comparisons of Self-reported Metrics between Real-world and VR Conditions; Error Bars Indicate Model-estimated 95% Confidence Intervals

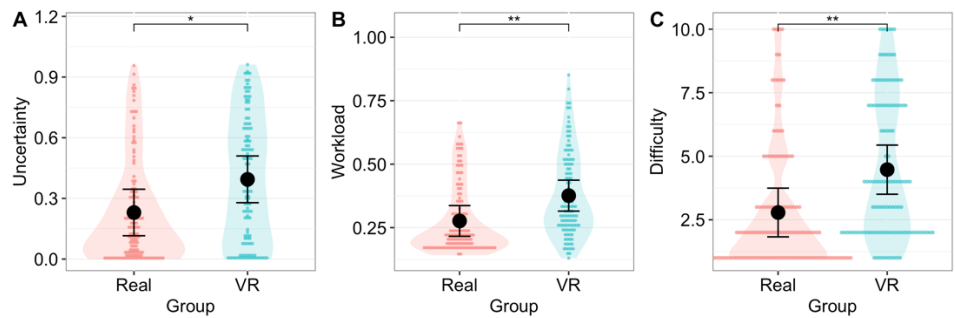

### 3.5. The Effect of Age (RQ4)

Results for the impact of age are presented in Table 4. We were not able to reject any of the null hypotheses, hence age was not found to have a significant interaction effect for any of the dependent variables. The Bayes Factors analysis strongly suggested ($BF_{10} < 10^{-1}$) the exclusion of age interaction effects in models for five of the dependent variables: Number of Mistakes, Backtracking, Uncertainty, Workload, and Directional Pointing; and it moderately suggested exclusion of age interaction effects for the Task Difficulty variable ($BF_{10} < 10^{-1/2}$). However, the Bayes Factors analysis was close to neutral ($10^{-1/2} < BF_{10} < 10^{1/2}$) for Sign-seen Frequency and Sign-seen Duration, and it moderately suggested the inclusion ($BF_{10} > 10^{1/2}$) of interaction effects of age in Distance Traveled and Task Duration.

**Table 4.** F-test Results and $BF_{10}$ of Age by Condition (Real-world vs. VR)

| Dependent Measure | Estimated Contrast [95% CI] | Denominator df | F | P | ω2 | Partial η2 | $BF_{10}$ |
|---|---|---|---|---|---|---|---|
| Distance Traveled | 14.77 [-48.29, 77.84] | 231.06 | 0.23 | 0.633 | <0.01 | <0.01 | 5.593 |
| Number of Mistakes | 0.04 [-1.05, 1.13] | 223.01 | 0.01 | 0.941 | <0.01 | <0.01 | 0.088 |
| Task Duration | 34.19 [-56.38, 124.75] | 32.99 | 0.59 | 0.447 | <0.01 | 0.02 | 9.624 |
| Backtracking | -0.10 [-0.99, 0.78] | 231.07 | 0.06 | 0.810 | <0.01 | <0.01 | 0.072 |
| Sign-seen Frequency | 1.44 [-7.01, 9.89] | 27.48 | 0.12 | 0.731 | <0.01 | <0.01 | 0.707 |
| Sign-seen Duration | -8.65 [-30.16, 12.86] | 30.23 | 0.67 | 0.419 | <0.01 | 0.02 | 2.380 |

| Uncertainty | -0.01 [-0.31, 0.29] | 32.05 | 0.00 | 0.958 | <0.01 | <0.01 | 0.023 |
| Workload | 0.05 [-0.07, 0.17] | 31.84 | 0.73 | 0.400 | <0.01 | 0.02 | 0.013 |
| Difficulty | 0.46 [-1.64, 2.56] | 31.76 | 0.20 | 0.659 | <0.01 | <0.01 | 0.183 |
| Directional Pointing | 0.01 [-0.18, 0.21] | 31.88 | 0.02 | 0.880 | <0.01 | <0.01 | 0.015 |

*Note:* Estimated Contrast: ({Old, VR} - {Old, Real}) - ({Young, VR} - {Young, Real}). F-tests with Satterthwaite's method, numerator df = 1. Denominator dfs are larger for some measures due to no random effects of participants from REML estimations.

### 3.6. Distribution of Uncertainty across the Floorplan (RQ5)

F-tests found no significant effects of real-world vs. VR conditions on uncertainty distribution. There were also no significant effects found for zone type (decision or non-decision zones) on uncertainty, and no significant effects for the interaction of the experiment conditions and zone types (Table 5). The spatial maps of uncertainty shown in Figure 7 indicated that higher uncertainty usually occurred in long corridors and at decision points. There was a generally consistent overall trend of lower uncertainty levels in the real-world condition, but this was inverted for some areas around stairwells and building entrances.

**Table 5.** F-Test Results for the Effects of Experimental Condition (Real-world vs. VR) and Zone Type (Decision vs. Non-decision), as Well as Their Interaction Effect

| Predictor | Denominator df | F | P | ω2 | Partial η2 |
|---|---|---|---|---|---|
| Condition (Real vs. VR) | 38.91 | 0.76 | 0.387 | <0.01 | 0.02 |
| Zone Type (Decision vs. Non-decision) | 24.49 | 1.56 | 0.223 | 0.02 | 0.06 |
| Condition by Zone Type | 622.50 | <0.01 | 0.991 | <0.01 | <0.01 |

*Note:* F-tests with Satterthwaite's method, numerator df = 1.

**Figure 7.** Spatial Distribution of Uncertainty (RE = Real Environment; VE = Virtual Environment; DIFF = Difference between RE and VE)

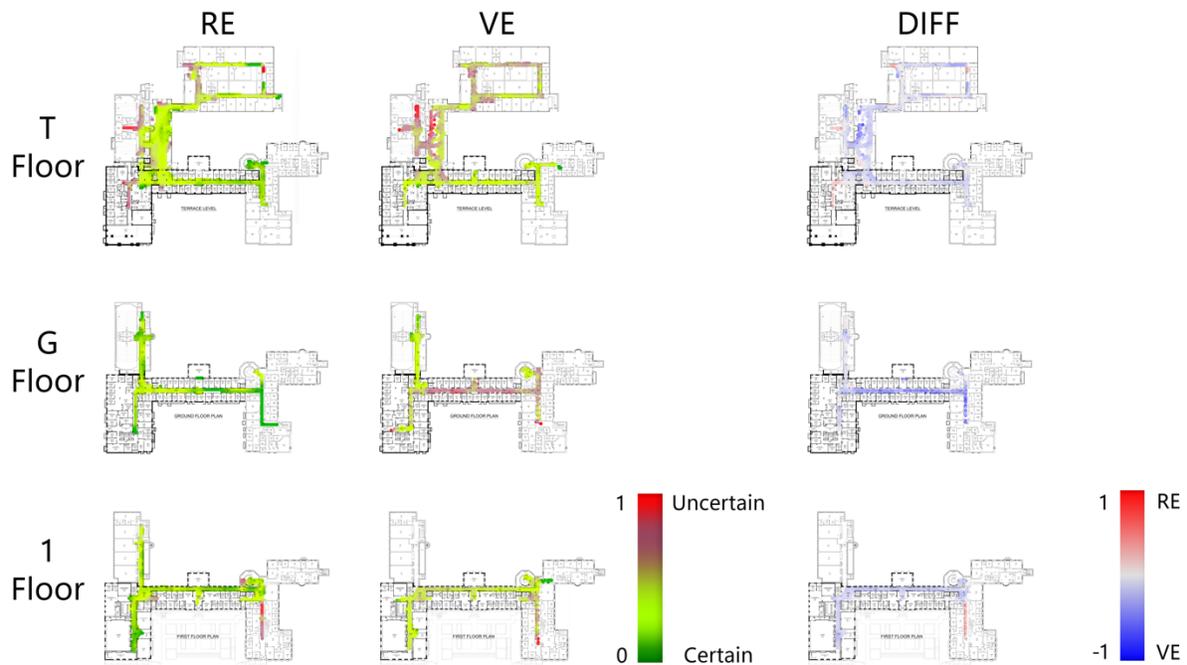

## 4. Discussion

The central aim of this study was to contribute to the literature on differences and similarities in navigation-related human responses in VR vs. real-world environments, with a particular focus on the responses of older adults. We identified significant VR/real-world response differences across every measure that we evaluated, and we did not find any interactions effect of age on these VR/real-world differences.

**Wayfinding Performance and Spatial Learning (RQ1).** The performance metrics used in the current study, including distance traversed, number of errors, and task duration, are very commonly deployed in human navigation research (Iftikhar et al., 2021). Our findings indicated marked variations in these measures between the real-world facility and the identical VR setting, which is incommensurate with previous research suggesting VR/real-world congruity in these measures (Ewart & Johnson, 2021; Lloyd et al., 2009; Skorupka, 2009). Our study indicated that participants in the VR environment traveled longer distances, made more mistakes, and took longer to complete wayfinding tasks compared to the real-world environment; furthermore, their subsequent cognitive maps appeared to be less accurate as measured by directional pointing error. These outcomes, we posit, may be attributable to the greater ease of exploration in the VR (requiring less physical effort), and possibly to a reduced senses of urgency or import in the VR task completion, leading to more exploratory or "playful" approaches. Some previous researchers have similarly suggested that the interactive aspect of VR, which can resemble gaming scenarios, may stimulate participants to explore more extensively and less urgently than in real-world settings (Kuliga et al., 2020; Skorupka, 2009). It may also be possible that the lack of fully nuanced environmental cues in the VR contributed to the reduced wayfinding performance, or that the use of joystick-derived movement contributed to disorientation and reduced cognitive mapping outcomes (Cherep et al., 2022; Sarupuri et al., 2017). Further study on participant motivations and wayfinding strategies would be needed to confirm such interpretations. Regardless of the explanation, these findings suggest that when VR is used to contrast varying design conditions the resulting data should be interpreted cautiously and should not be directly translated into real-world conclusions without further justification.

**Wayfinding Search Behavior (RQ2).** The process of human navigation in built environments involves an array of specific information-seeking behaviors, including backtracking, sign-seen frequency, and duration of examining signs. Our study identified significant discrepancies in all three of these factors between the real-world facility and the identical VR environment, with heightened information-seeking behaviors found in the VR. The reasons for these behavioral differences are not entirely clear, but they may be related to the greater perceived workload and uncertainty that participants associated with the VR environment, as discussed in the following paragraph.

**Uncertainty, Cognitive Load, and Perceived Difficulty (RQ3).** Our study identified significant disparities in self-reported psychological response outcomes between the VR and real environments. Participants in the VR conditions reported higher levels of perceived uncertainty, workload, and task difficulty. These results may in part be intrinsic to the VR experience; but we suspect that the novelty of the VR technology for study participants also played a role. While robust data on prior VR use was not collected in the current study, it seems safe to say that the participants had greater previous experience with the real-world wayfinding than they had with VR; and our anecdotal observations indicated that for many participants this was their very first time using VR. The findings of greater perceived uncertainty, cognitive load, and difficulty in the VR setting may be regarded as a potential explanatory factor for the differences that we found in wayfinding behavior and performance, as information-seeking behaviors motivated by these perceptions could lead to greater exploration, more time spent looking for signs and other environmental cues, and an associated reduction in performance outcomes. This interpretation is supported by prior literature that has demonstrated an influence of perceived uncertainty on the selection of actions during wayfinding (Hirsh et al., 2012; Savino et al., 2019). Even apparently unrelated behaviors such as backtracking could potentially be explained by the uncertainty reduction theory, if greater perceived uncertainty makes it more likely for participants to abandon a chosen path in favor of exploring new directions. The interplay between uncertainty, cognitive load, information-seeking behavior, and spatial exploration warrants further study to elucidate these dynamics fully.

**Effects of Participant Age (RQ4).** The study was also motivated by the need to understand the interaction effects of age on spatial wayfinding in VR versus real environments. This interest stems from the well-documented decline in spatial navigation abilities with age and the potential of VR for applications in aging—such as early detection and intervention for conditions like Alzheimer's disease. Our analysis of age interaction with VR and real-world conditions in metrics such as wayfinding performance, perceived task difficulty, and spatial learning yielded non-significant results with small effect sizes. This suggests minimal age-based moderation on the discrepancies between VR and real-world spatial navigation tasks. Our results indicate that older adults adapt to VR environments comparably to younger counterparts, with their spatial navigation performance mirroring real-world situations. This aligns with prior research underscoring the ability of older adults to adeptly navigate VR environments for spatial tasks (Kalantari, Bill Xu, et al., 2022; Kimura et al., 2017b; Molina et al., 2014). It hints that cognitive strategies for spatial navigation remain consistent across ages, irrespective of the setting (VR or real). Nevertheless, caution is advised. The lack of significant interaction may be attributed to our study's constraints, such as its power or sensitivity in measurements. Future studies should delve deeper, examining the influence of distinct VR navigation interfaces or other age-specific cognitive or physical factors on navigation disparities between VR and the real world.

**Spatial Distribution of Uncertainty (RQ5).** We did not find a significant difference in the spatial distribution of uncertainty between the VR and real-world conditions. In other words, the significantly greater uncertainty found in the VR condition appeared to be distributed throughout the floorplan, rather than associated with any specific geographical zones. This finding suggests that participants used comparable strategies and cues for navigation in both settings. Our analysis also revealed that higher uncertainty was clustered around specific areas of the facility (identical in both real-world and VR), which primarily included route-decision points and long hallways. This finding was expected, and it is commensurate with prior research that has highlighted the importance of environmental factors in spatial navigation (Devlin, 2014). The correspondence of high-uncertainty zones in the real-world and VR indicates that the relevant environmental features were similar in both contexts, although the extent of participants' responses to these features was significantly different.

Our study enriches the understanding of human behavior in the burgeoning realm of VR, focusing on spatial navigation. We determined that although VR retains some ecological validity, it elicits distinct behavioral and cognitive outcomes compared to real-world environments. These results challenge certain prevailing notions in current literature, highlighting the imperative for judicious interpretation and application of VR in spatial studies. We identified pronounced differences in metrics such as wayfinding performance, spatial learning, perceived uncertainty, cognitive workload, and task difficulty between VR and real-world scenarios. Notably, participants in VR exhibited heightened exploration, errors, and uncertainty, likely owing to VR's immersive nature and the disorientation from teleportation mechanics. Yet, there was a remarkable parallel in how uncertainty was spatially distributed in both VR and real-world contexts, hinting at the adoption of akin navigation strategies. On the subject of age, while it is recognized that spatial navigation abilities diminish over time, our findings debunked age as a significant modulator of VR versus real-world navigation disparities. This means that older adults acclimatized to VR almost as seamlessly as their younger counterparts.

For scholars venturing into VR for spatial navigation studies, it is crucial to acknowledge VR's unique intricacies, including its propensity to induce heightened exploration and cognitive demand. Our results demonstrate that while participants traverse VR landscapes in ways that echo real-world navigation, marked differences still prevail. For those investigating the nexus of aging and spatial navigation, VR emerges as a promising tool, with age not drastically altering outcomes. Nevertheless, the exact VR interfaces, tasks, and other underlying factors warrant deeper exploration to fathom this interaction fully. In sum, our research bridges gaps in VR spatial navigation literature, offers a roadmap for subsequent studies, and underlines the utility of VR in researching and aiding spatial navigation across diverse age groups. It sets the stage for more in-depth inquiries into VR's multifaceted spatial navigation landscape and the design of universally accessible VR systems.

**Limitations and Future Directions.** The findings of the current study provide important evidence about the existence of real-world/VR discrepancies in wayfinding responses. However, the study does have some limitations. First, the use of a convenience sampling method means that our participants may not be entirely representative of the wider population, and the findings may be subject to selection bias. Beyond the consideration of age-related effects, we did not evaluate the potential effects of other demographic variables that are potentially salient to wayfinding outcomes or responses to VR, such as gender, income, and geographic background.

Second, the research was conducted in a single facility (duplicated in VR), which might limit the generalizability of the findings to other settings. The architectural structure and design of the chosen location, as well as the signage used, could have potentially influenced the observed wayfinding outcomes. Finally, while we endeavored to ensure that the VR context closely mirrored the real-world facility, some furnishing elements may have differed slightly in positioning and appearance (Figure 1), which could have affected the wayfinding outcomes.

In addition to improving upon these limitations, we recommend that future studies in this area incorporate measures of human-response variables that do not rely on self-reporting, for example by using heart rate or skin conductance to capture the affective responses more objectively. With the rapid advancement in VR technology, future studies could further refine the VR environments to better emulate real-world settings, for example by introducing sound recordings and dynamic elements such as moving objects and crowds corresponding to the real-world conditions. Finally, future research should consider the interplay among various factors influencing spatial abilities and wayfinding behaviors, such as models that simultaneously reflect gender and age, rather than treating them as autonomous variables. Such an approach could provide a more nuanced understanding of how different factors work together to shape individuals' performance and behaviors in real-world and VR environments.